\documentclass[aps,pra,twocolumn,groupedaddress,floatfix]{revtex4-1}
\usepackage{graphicx}
\usepackage{amsmath}
\begin{document}

\title{Spin-Imbalanced Quasi-Two-Dimensional Fermi Gases}

\author{W.~Ong$^{1,2}$, C.-Y.~Cheng$^{1,2}$, I.~Arakelyan$^{1}$, and J.~E.~Thomas$^1$}

\affiliation{$^{1}$Department of  Physics, North Carolina State University, Raleigh, NC 27695, USA}
\affiliation{$^{2}$Department of Physics, Duke University, Durham, NC 27708, USA}

\date{\today}

\begin{abstract}
We measure the density profiles for a Fermi gas of $^6$Li containing $N_1$ spin-up  atoms and $N_2$ spin-down atoms, confined in a quasi-two-dimensional geometry. The spatial profiles are measured as a function of spin-imbalance $N_2/N_1$ and interaction strength, which is controlled by means of a collisional (Feshbach) resonance.  The measured cloud radii and central densities are in disagreement with mean-field Bardeen-Cooper-Schrieffer (BCS) theory for a true two-dimensional system. We find that the data for normal-fluid mixtures are reasonably well fit by a simple two-dimensional polaron model of the free energy. Not predicted by the model is a phase transition to a spin-balanced central core, which is observed above a critical value of $N_2/N_1$.
Our observations provide important benchmarks for predictions of the phase structure of quasi-two-dimensional Fermi gases.
\end{abstract}

\maketitle

Layered strongly correlated systems play important roles in the quest for high temperature superconductors. In high-transition temperature copper oxide and organic compounds, electrons are confined in a quasi-two-dimensional geometry, creating complex, strongly interacting many-body systems, for which the phase diagrams are not well understood~\cite{Norman08042011}. The basic underlying mechanism for superconductivity, pairing of fermions, can be disrupted by an unequal number of pairing species when the Fermi surfaces of the two spin components are mismatched, leading to exotic superconductivity in which pairs acquire finite momentum~\cite{MayafreeFFLO}. Such spin-imbalanced Fermi mixtures also can contain polarons, quasiparticles formed by mobile impurities in a fermionic bath.  Ultracold atomic Fermi gases provide a new platform for emulation of these systems, with precise experimental control~\cite{PhysRevLett.102.230402, PhysRevLett.105.030404, Feld2011, PhysRevLett.108.045302, Koschorreck2012, JochimPairCondensation}.

Previous studies of pairing in spin-imbalanced three-dimensional (3D)~\cite{PhysRevLett.97.030401,Partridge27012006,Nature451-2008} and one-dimensional (1D)~\cite{Liao2010} Fermi gases revealed phase separation. In 3D, a spin-balanced, fully-paired, superfluid core is surrounded by an imbalanced normal fluid shell, followed by a fully polarized shell, a structure successfully described by an elegant polaron model~\cite{0034-4885-73-11-112401}. For measurements in 1D imbalanced mixtures, the behavior is reversed: A balanced phase appears outside a spin-imbalanced core, in agreement with a mean field model.

\begin{figure}[t]
\includegraphics[width=2.25in]{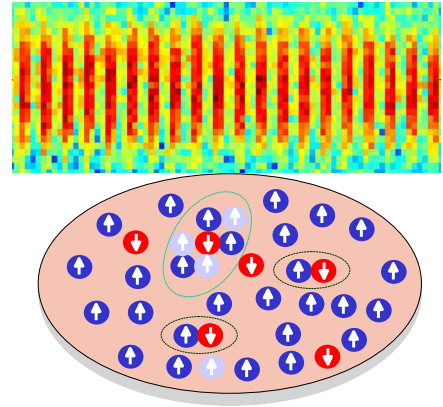}
\caption{Top: Side image of layered pancake-shaped atom clouds, separated by 5.3 $\mu$m in a CO$_2$ laser standing-wave trap. Bottom: In each pancake, confinement causes majority spins (blue-up arrow) and minority spins (red-down arrow) to pair, producing bound dimers. Polarons form when minority atoms scatter in the Fermi sea of the majority atoms and become surrounded by a cloud of particle-hole pairs (dark-blue-light-blue). Tightly bound dimers also scatter, forming dressed dimers.\label{fig:polaron-dimer}}
\end{figure}

A natural question is how the phase diagram of a quasi-two-dimensional cloud, containing a spin-imbalanced Fermi gas, differs from those measured in one and three dimensions. Does phase-separation occur? If so, what separates? Unlike a 3D gas in free space, a two-dimensional (2D) gas naturally contains bound dimers~\cite{PhysRevLett.62.981,PetrovShlyapnikov2D}. The binding energy of these dimers, $E_b\geq 0$, sets the natural scale of length for scattering interactions in 2D systems. 2D-Polarons~\cite{PhysRevA.83.021603} may be important for a quasi-2D Fermi gas~\cite{PhysRevLett.108.235302}. The phase diagram for imbalanced mixtures in this regime is therefore likely to be very rich~\cite{PhysRevB.89.014507,PhysRevB.90.214503}, involving the interplay and phase separation of several components, including dimer gases, polaron gases and spin-imbalanced normal fluids, as shown in~Fig.~\ref{fig:polaron-dimer}. Exotic components with spatially modulated superfluids (Fulde-Ferrell-Larkin-Ovchinnikov states), and vortex-anti-vortex pairs (Berezinskii-Kosterlitz-Thouless states) also have been predicted for 2D and quasi-2D Fermi gases~\cite{PhysRevLett.95.170407,PhysRevLett.96.040404,PhysRevA.78.043617,PhysRevA.79.053637,
PhysRevLett.106.110403,PhysRevLett.112.135302,
PhysRevB.89.014507,Sheehy2DFFLO}. The dimensionality of a single layer in Fig.~\ref{fig:polaron-dimer} is determined by the ratio of the transverse Fermi energy $E_{F}$  to the energy level spacing $h\nu_z$ in the tightly confined z-direction. The system is two-dimensional if $E_F/h\nu_z<<1$ or three dimensional if $E_F/h\nu_z>>1$.

\begin{figure*}[htb]
\includegraphics[width=\textwidth]{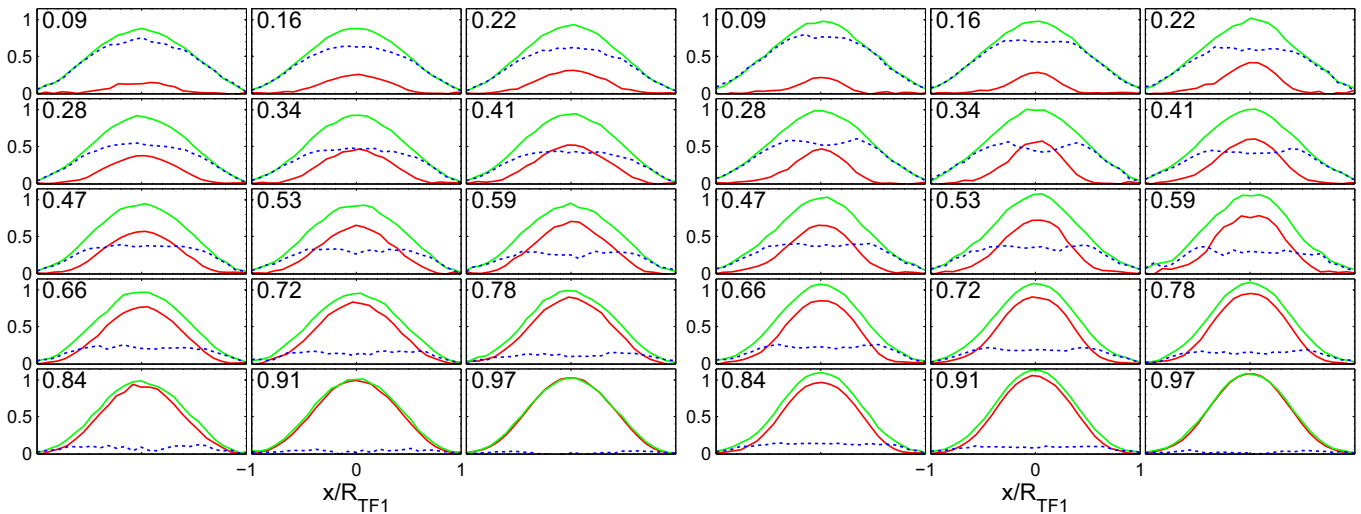}
\caption{Measured column density profiles in units of $N_1/R_{TF1}$ at 832 G, for $E_F/E_b=6.6$ (left panel) and at 775 G, for $E_F/E_b=0.75$ (right panel) versus $N_2/N_1$. Green: 1-Majority; Red: 2-Minority. Blue-dashed: Column density difference. Each profile is labeled by its $N_2/N_1$ range. For the density difference, the flat centre and two peaks at the edges are consistent with a fully paired core of the corresponding 2D density profiles. These features are more prominent for the higher interaction strength (right panel).\label{fig:profiles}}
\end{figure*}

We report measurements of the spatial profiles for spin-imbalanced mixtures in the intermediate quasi-two-dimensional regime~\cite{PhysRevLett.108.235302}, where $E_F/(h\nu_z)\simeq 1$. This regime is of great interest, as the onset of a superfluid phase is predicted~\cite{PhysRevB.90.214503} to occur at a higher critical temperature than for a true 2D system. Control of the relative spin-population permits precision studies of the phase diagram for these quasi-2D-gases, which has been the topic of intense theoretical study~\cite{PhysRevLett.62.981,PhysRevLett.95.170407,PhysRevB.90.214503,
PhysRevLett.96.040404,PhysRevLett.96.110403,PhysRevA.78.043617,PhysRevA.79.053637,
PhysRevLett.106.110403,PhysRevLett.112.135302,PhysRevB.89.014507,Sheehy2DFFLO}.

\begin{figure*}[t]
\includegraphics[width=5.8 in]{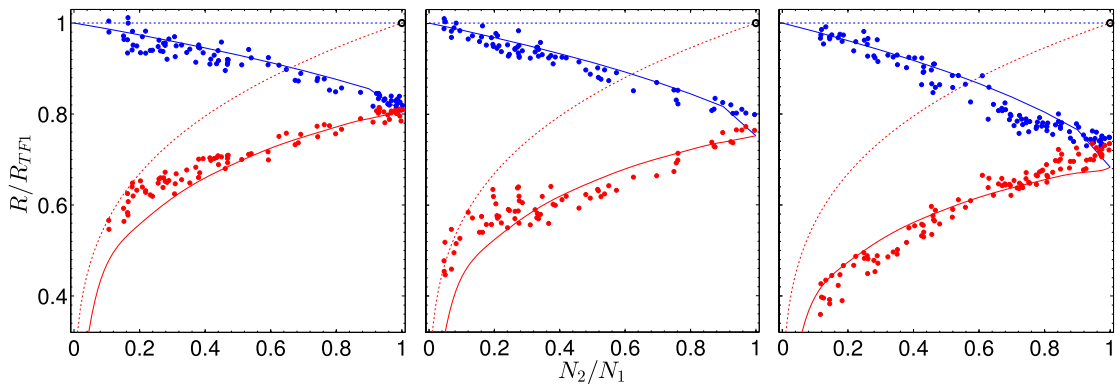}
\caption{Majority radii (upper-blue) and minority radii (lower-red) in units of the Thomas-Fermi radius for the majority for $E_{F}/E_b=6.6$ (left panel), $E_{F}/E_b=2.1$ (middle panel) and $E_{F}/E_b=0.75$ (right panel). Dots: Data; Solid lines: 2D polaron model; Dashed lines: Ideal Fermi gas prediction; Black circle upper right: 2D-BCS theory for a balanced mixture.\label{fig:radii}}
\end{figure*}

We investigate density distributions of imbalanced quasi-2D gases by direct imaging. Measured column density profiles for various interaction strengths, $E_{F}/E_b$  are  shown in  Fig.~\ref{fig:profiles}. From the column densities, we extract the radii and the central 2D densities for each state, as a function of $N_2/N_1$, where $N_1$ is the number of majority atoms and $N_2$ the number of minority atoms. For simplicity, the cut-off radii are determined by fitting the measured column density with the spatial profile for an ideal 2D gas, $n_{1D}(x)=n_0(1-x^2/R^2)^{3/2}$. Fig.~\ref{fig:radii} shows the cloud radii for the majority (blue dots) and minority (red dots), for $E_{F}/E_b=6.6,\,\,2.2$, and $0.75$.  Both radii are given in units of the Thomas-Fermi radius $R_{TF1}$ for the majority, to clearly demonstrate the deviation from the predictions for an ideal Fermi gas, which is shown for comparison as the blue-dashed and red-dashed curves. For the nearly polarized clouds, where $N_2/N_1=0.1$, the measured majority radii approach the ideal gas limit. As the $N_2/N_1$ is increased, the measured radii of both species are significantly affected by attractive interactions between the two spin components.

To consider many-body interactions, we first compare the measured cloud radii for the balanced mixture, $N_2/N_1=1$, with BCS theory predictions for a true 2D Fermi gas~\cite{PhysRevLett.62.981,1742-5468-2012-10-P10019}, which shows $\epsilon_{F}=\mu+E_b/2$. This yields profiles identical to those of an ideal gas~\cite{Supplement}, leading to $R/R_{TF1}=1$ for both spin states (black circle Fig.~\ref{fig:radii}), in disagreement with the measured radii, which are much smaller.

Now we compare the data in Fig.~\ref{fig:radii} to a simple 2D polaron model, which is briefly summarized here and described in detail in the Supplementary Information~\cite{Supplement}.
At zero temperature, the free energy density $f$ is equal to the energy density. For an imbalanced mixture, with $N_2<<N_1$, we assume the 2D energy density is
\begin{equation}
f=\frac{1}{2}\,n_{1}\,\epsilon_{F1}+\frac{1}{2}\,n_{2}\,\epsilon_{F2}+n_{2}\,E_p(2).
\label{eq:freeenergyimbal}
\end{equation}
Here, the first two terms are the energy density for a noninteracting gas and the last term is the energy density for minority polarons in state $2$, which arises from scattering in the bath of majority atoms in state $1$; $n_{1,2}$ and $\epsilon_{F1,2}$ are the corresponding densities and local Fermi energies. The 2-polaron energy per particle $E_p(2)\equiv y_m(q_1)\,\epsilon_{F1}$, where $q_1\equiv \epsilon_{F1}/E_b$.  The function $y_m(q_1)$ is derived for a 2D gas in Ref.~\cite{PhysRevLett.108.235302}. For simplicity, we use an analytic approximation~\cite{PhysRevA.84.033607}, $y_m(q_1)=-2/\log(1+2\,q_1)$. From Eq.~\ref{eq:freeenergyimbal}, we directly obtain the local chemical potentials, $\mu_1=\partial f/\partial n_1$ and $\mu_2=\partial f/\partial n_2$ and the corresponding local 2D pressure $p=n_{1}\,\mu_{1}+n_{2}\,\mu_{2}-f$. The chemical potentials determine the spatial profiles in the trap.

The polaron model predictions for $R_1$ and $R_2$ are shown as the upper (blue) and lower (red) solid curves in Fig.~\ref{fig:radii}. Although the model is strictly valid only for small $N_2/N_1$, we display the predictions based on Eq.~\ref{eq:freeenergyimbal} for the imbalanced gas for $N_2/N_1=0$ up to $N_2/N_1=0.9$. For $N_2/N_1=1$, we show the predictions for the balanced mixture, which employs a spin-symmetrized free energy density~\cite{Supplement}.

\begin{figure}[htb]
\includegraphics[width=3.25in]{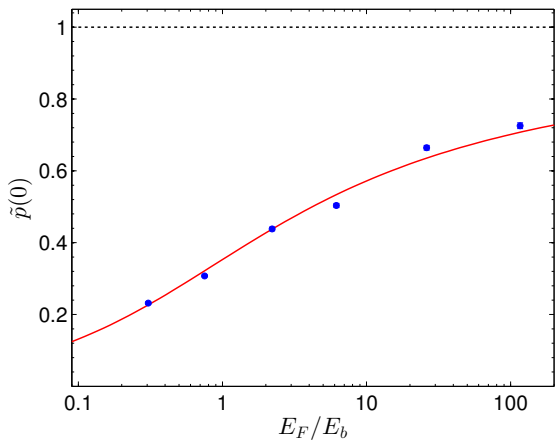}
\caption{Reduced 2D pressure at the trap center versus $E_{F}/E_b$ for the balanced gas. Dots: Experiment; Solid red curve: Prediction based on the polaron model for the balanced gas; Dashed line: Prediction of 2D-BCS theory.\label{fig:pressure}}
\end{figure}

The central pressure for the balanced gas ($N_2=N_1$) is determined by the 2D central density $n(0)$, which is  directly obtained from the measured central column density $n_{1D}(0)$. As discussed in the supplemental material~\cite{Supplement}, $p\propto 1/n(0)^2$ and $n(0)\propto n_{1D}^2(0)/N_1$. From this, we obtain the 2D pressure at the trap center in units of the ideal Fermi gas pressure for the same density, $\tilde{p}$, Fig.~\ref{fig:pressure}. The red solid curve shows the 2D polaron model prediction, for the same trap frequency $\omega_\perp$ as used to determine $R_{TF1}$ in the cloud profile measurements, with no other adjustable parameters.
For comparison, using the 2D-BCS theory prediction~\cite{PhysRevLett.62.981,1742-5468-2012-10-P10019}, where $\epsilon_{F}=\mu+E_b/2$, the Gibbs-Duhem relation requires  $\tilde{p}=1$ for all $E_F/E_b$, in contrast to the measurements~\cite{Supplement}.
\begin{figure*}[t!]
\includegraphics[width=6.0 in]{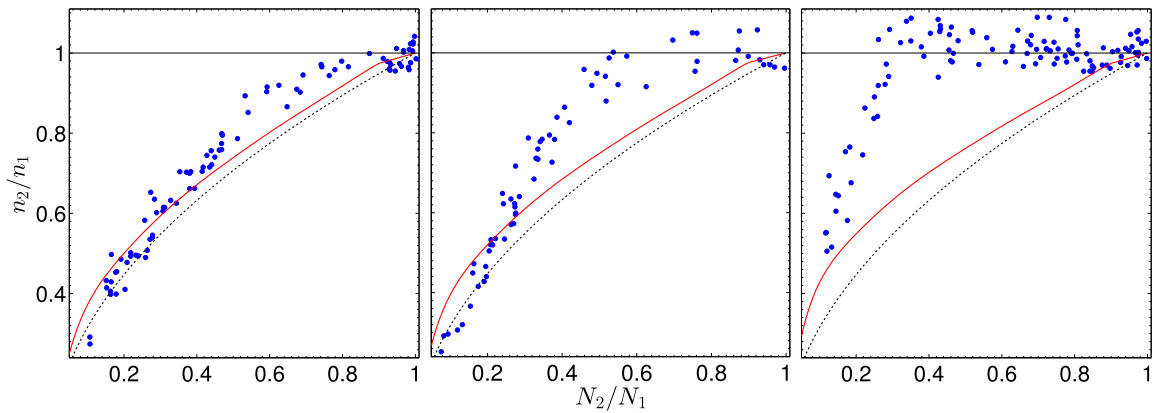}
\caption{Ratio of minority to majority 2D central densities for $E_{F}/E_b=6.6$ (left panel), $E_{F}/E_b=2.1$ (middle panel) and for the strongest interactions $E_{F}/E_b=0.75$ (right panel). Blue dots: Data; Red solid line: 2D polaron model; Dashed black line: Ideal Fermi gas prediction. Stronger interactions balance the central densities over an extended range of imbalance, in clear disagreement with the polaron model.\label{fig:densityratio}}
\end{figure*}

We have also measured the central density ratio $n_{2}/n_{1}$ of the 2D gas as a function of $N_2/N_1$. First, we fit a Thomas-Fermi 1D profile to each column density, from which we find the corresponding 2D densities as described above. Also, we employ an inverse Abel transformation of the column densities to extract the peak 2D densities. Both methods yield similar results within 5\%. We show the density ratios for three interaction strengths in Fig.~\ref{fig:densityratio}. The agreement with the polaron model is reasonably good  at 832 G, where $E_{F}/E_b=6.6$. However, as the interaction strength is increased to $E_{F}/E_b=0.75$ by increasing the dimer binding energy at $775$ G, the 2D central densities abruptly become balanced above a critical ratio $N_2/N_1$, right panel Fig.~\ref{fig:densityratio}. To ensure that the densities are balanced not just at the centre, but over an extended range, we examine the measured column density profiles in Fig.~\ref{fig:profiles}. The apparent presence of two peaks at the edges in the column density difference versus $x$ is consistent with the $y$-integrated 2D shell structure of a balanced core surrounded by an unpaired majority fraction~\cite{Supplement}. Note that double integration of the 3D shell structure gives rise to flat top distributions~\cite{PhysRevLett.97.030401}. Equal densities for any imbalance are not predicted by the 2D polaron model, as the pressure determined for the imbalanced gas from Eq.~\ref{eq:freeenergyimbal} is always greater than or equal to the pressure for the balanced gas, contrary to the 3D case where crossing of the two pressures determines the critical polarization for the phase separation~\cite{0034-4885-73-11-112401}. This is not unexpected as the simple polaron model with the analytic approximation for the polaron energy overestimates the magnitude of $E_p$ on the molecular side of the Feshbach resonance ($B<832$ G) and does not include the effective mass or molecular repulsion energy~\cite{PhysRevA.87.033616}.

In conclusion, the 2D polaron model explains much of the behavior of the spin-imbalanced normal fluid mixtures. However, more precise calculations of the pressures for the balanced and imbalanced components are needed to explain the observed phase separation and critical spin-imbalance. Our measurements will serve as a test for predicted phase diagrams, which will help to reveal the structure of a quasi-2D Fermi gas.

This research is supported by the Physics Division of the Army Research Office (Many-body physics in two-dimensional Fermi gases),  and the Division of Materials Science and Engineering, the Office of Basic Energy Sciences, Office of Science, U.S.
Department of Energy (Thermodynamics in strongly correlated Fermi gases). Additional support
has been provided by the Physics Division of the National Science Foundation and the Air Force Office of Scientific Research.   We thank M. Bluhm, T. Sch\"{a}fer, and D. Lee, North Carolina State University, for stimulating conversations.

%

\widetext
\appendix
\section{Supplementary Information}
\label{sec:supplement}

In this supplementary material, we first discuss the experimental methods. Then we show that BCS theory for a true 2D Fermi gas~\cite{PhysRevLett.62.981,1742-5468-2012-10-P10019} fails to account for the measured density profiles of a balanced mixture. Finally, we present a simple 2D polaron model for the density profiles and the pressure of a two-component 2D Fermi gas. In our simplified treatment, we ignore the effective mass and do not include a molecular dimer pressure~\cite{1367-2630-13-11-113032} or molecular dimers and dimer-polarons~\cite{PhysRevA.87.033616}. We will separately consider both spin-imbalanced and spin-balanced clouds. For the balanced gas, we show that the predicted pressure is in good agreement with the ground-state pressure that we measure from the central density and in reasonable agreement with the pressure  measured for 2D Bose and Fermi gases over a broader range of interaction strengths near the 3D Feshbach resonance in $^6$Li~\cite{PhysRevLett.112.045301}.

\subsection{Experimental Methods}
In the experiments, a 50-50 mixture of the two lowest hyperfine states of $^6$Li, denoted $1$ and $2$, is confined in a single-beam CO$_2$ laser trap and initially cooled by forced evaporation near the Feshbach resonance at 832 G~\cite{O'Hara13122002}. During the forced evaporation, the trapping laser intensity is gradually reduced and a retro-reflected trapping laser beam is smoothly turned on, creating a standing wave at low trapping beam intensity. The bias magnetic field is then slowly swept to 1200 G, where the cloud is weakly interacting, and a frequency-swept, radio-frequency pulse is applied, which is resonant with a transition from state $2$ to a higher lying hyperfine state $3$. This depletes the population of state $2$. Varying the pulse amplitude controls the population imbalance between states $1$ and $2$. Atoms in state $3$ are subsequently removed by an optical pulse on a time scale that is fast compared to the time for three-body inelastic collision to occur. The bias magnetic field is then swept to the field of interest, which sets the dimer binding energy and thus the interaction strength $E_{F}/E_b$. The standing wave trap is further lowered to cool the gas to an energy near the ground state. Using this method, the number of atoms in the majority state $N_1\simeq 800$ is held approximately constant as $N_2$ is varied in the experiments. In addition, the ratio $N_2/N_1$ is nearly the same for each two-dimensional trap site in the standing wave, as the initial 50-50 mixture  is trapped in the standing wave before the radio-frequency pulse is applied. All experiments are conducted with the same trap frequencies, $\omega_z = 2\pi\times9$ kHz along the tight confinement direction and $\omega_\bot = 2\pi\times440$ Hz in the transverse direction. Thus, the transverse Fermi energy of the majority component is nominally the same for all experiments.

The quoted $\omega_\perp$ is $8$\% higher than that measured at the peak of a parametric resonance at $1200$ G in our shallow trap, where we expect anharmonicity to lower the resonance frequency. Indeed, the measured resonance frequency yields a Thomas-Fermi radius $R_{TF1}$ for an ideal gas that is larger than the measured majority radius $R_1$ for $N_2/N_1\rightarrow 0$, which is not physical. We therefore increase the frequency $\omega_\perp$ so that the majority radius $R_1$ measured at $1200$ G approaches $R_{TF1}$ as $N_2/N_1\rightarrow 0$.  Without further adjustment, this frequency is used to compute the Thomas-Fermi radius for all reported measurements. Typical scales are the following: trap potential depth $U_0 = 3.3$  $\mu$K, ideal gas Fermi energy $E_{F1} = 0.85$ $\mu$K, Thomas-Fermi radius $R_{TF1} = 17$ $\mu$m, axial level spacing $h\nu_z = 0.43$ $\mu$K; at 775 G, $E_b = 1.15$ $\mu$K, and the transverse energy $m\omega^2_\bot\left\langle x\right\rangle^2 = 0.16$ $\mu$K. $E_b$ is determined as in ref. 15 including the small transverse confinement $\omega_\perp/\omega_z=0.05$, which increases $E_b$. The effect on the polaron model is small at and below the Feshbach resonance, due to the logarithmic dependence on $E_F/E_b$. However, for very weak binding, at $1200$ G, the effect is large, and $E_F/E_b=116$ for $\omega_\perp/\omega_z=0.05$, while $E_F/E_b\simeq 10^4$ for $\omega_\perp=0$.

The spatial profiles of both spin states $1$ and $2$ are measured for the same cloud and determined after a short time of flight by absorption imaging with a CCD camera, using sequential imaging beam pulses, each resonant with one of the states. In our CO$_2$ laser standing wave trap, the transverse profiles are photographed essentially \textit{in situ} as we use an expansion time $t_{exp} = 30-40$ $\mu$s $<< 1/\omega_\bot$. On the contrary, as $t_{exp} \simeq 1/\omega_z$, the individual clouds expand significantly in the tightly confined direction, which lowers the optical density and reduces systematic effects~\cite{PhysRevLett.112.045301}. Our imaging pulses are 5 $\mu$s long, separated by 5 $\mu$s, with intensities just below the saturation intensity. The column density profiles for each image are obtained by averaging over 20 central pancake clouds along the tight confinement direction. For Fig.~\ref{fig:profiles}, we show column density profiles averaged over a range of imbalance $N_2/N_1$ in steps of 0.06. Each profile is labeled by the central value of the corresponding range.

As imaging of the first spin state affects the image of the second, we rescale the amplitude of the second density profile by a maximum of 11\% to match that of the first in the case of balanced clouds. Cut-off radii and central 2D densities are extracted using ideal gas Thomas-Fermi distributions for each image of 20 pancake clouds and displayed in Figs.~\ref{fig:radii} and ~\ref{fig:densityratio}. We also use the inverse Abel transformation to obtain model-independent values of 2D central densities consistent with those obtained from the Thomas-Fermi distributions.

\subsection{Density profiles for BCS theory in 2D}
We first consider predictions of BCS theory for a true 2D Fermi gas~\cite{PhysRevLett.62.981,1742-5468-2012-10-P10019}. The theory shows that the local Fermi energy for a balanced superfluid mixture is $\epsilon_{F1}=\mu+E_b/2$, where  $\epsilon_{F1}=\alpha\,n_{1}$ and $\mu$ is the local chemical potential. Here, $n_{1}=n_{2}$ is the local density and $\alpha\equiv 2\pi\hbar^2/m$, with $m$ the atom mass. For a trapping potential $U_{trap}(\rho)$, $\mu=\mu_0-U_{trap}(\rho)$, with $\mu_0$ the chemical potential at the trap center and $\rho$ the transverse radius. Setting $\epsilon_{F1}(0)=\mu_0+E_b/2$, it is apparent that the predicted density profile is of the Thomas-Fermi form. For a harmonic trap, $U_{trap}(\rho)=m\omega_\perp^2\rho^2/2$, one obtains $n_{1}(\rho)=n_{1}(0)(1-\rho^2/R_{TF1}^2)$, where $n_{1}(0)=\epsilon_{F1}(0)/\alpha$. Normalization gives $\epsilon_{F1}(0)\equiv E_{F}=\hbar\omega_\perp\sqrt{2N_1}$,  the Fermi energy of an ideal Fermi gas at the trap center. The radius at which the density vanishes is the corresponding Thomas-Fermi radius, $R_{TF1}=\sqrt{2E_{F}/(m\omega_\perp^2)}$.
For the spin-balanced cloud, $N_2/N_1= 1$, the 2D-BCS prediction is then identical to that of an ideal gas, $R/R_{TF1}=1$ for both spin states, in disagreement with the measured radii, which are much smaller. We conclude that the BCS mean-field theory (MFT) is, as expected~\cite{PhysRevLett.62.981}, not adequate for describing our system with intermediate coupling, where interparticle spacing is comparable with the pair size. This sets our experiment apart from previously studied nearly 2D systems that produced a good agreement with the MFT predictions~\cite{doi:10.1146/annurev-conmatphys-031113-133829}.

\subsection{2D-Polaron Model: Spin-imbalanced mixtures.}
\label{Sec:freenergy}
We begin by assuming a simple form for the free-energy of a spin-imbalanced mixture, with $1$ the majority component and $2$ the minority, i.e., $N_2<N_1$ for the total number of atoms in each state. At zero temperature, the free energy density is equal to the energy density. For the imbalanced mixture, we take
\begin{equation}
f=\frac{1}{2}\,n_{1}\,\epsilon_{F1}+\frac{1}{2}\,n_{2}\,\epsilon_{F2}+n_{2}\,E_p(2).
\label{eq:Sfreeenergyimbal}
\end{equation}
Here, the first two terms are the energy density for a 2D noninteracting gas. The last term is the energy density for polarons in state $2$. The polaron energy $E_p(2)$ arises from scattering of state $2$ atoms from the Fermi sea of atoms in state $1$, immersing each state $2$ atom in cloud of particle-hole pairs of state $1$~\cite{PhysRevA.74.063628}.

The polaron energy $E_p(2)$ is negative and proportional to the local Fermi energy of the majority component, $\epsilon_{F1}=\alpha\,n_{1}$, where $\alpha\equiv 2\pi\hbar^2/m$ and $n_{1}$ is the number of atoms per unit area in state $1$,
\begin{equation}
E_p(2)=y_m(q_1)\,\epsilon_{F1}.
\label{eq:S1.6a}
\end{equation}
For the 3D problem at resonance, where $y_m=-0.6$~\cite{PhysRevA.74.063628}, the same method, with $1/2\rightarrow 3/5$ in Eq.~\ref{eq:Sfreeenergyimbal}, yields the Fermi liquid equation used in ref.~\cite{PhysRevA.77.041603} for the normal imbalanced mixture, with $m^*=1$. In ref.~\cite{PhysRevLett.108.235302}, $E_p(2)$ is obtained for a 2D gas as a function of $-\log q_1$, where $q_1\equiv \epsilon_{F1}/E_b$ and $E_b$ is the binding energy of $1-2$ dimers pairs in the 2D trap.  To simplify the treatment, we use instead an analytic approximation for $E_p(2)$ due to Klawunn and Ricati~\cite{PhysRevA.84.033607}, which interpolates between polaron behavior in the BCS regime and molecular behavior in the BEC regime,
\begin{equation}
y_m(q_1)=\frac{-2}{\log(1+2q_1)}.
\label{eq:S1.6}
\end{equation}
\noindent The analytic approximation has the correct behavior for $\epsilon_{F1}<<E_b$, where $q_1<<1$. For this case, we see that Eq.~\ref{eq:S1.6} yields $E_p(2)\rightarrow -E_b-\epsilon_{F1}$ as it should~\cite{PhysRevA.87.033616}. For the balanced gas, this result also yields the correct chemical potential per single atom for $\epsilon_{F1}<<E_b$ , as discussed below, see Fig.~\ref{fig:balchempot}.

From Eq.~\ref{eq:Sfreeenergyimbal}, we directly obtain the chemical potentials, $\mu_i=\partial f/\partial n_i$. For an atom in state $2$,
\begin{equation}
\mu_{2}=\epsilon_{F1}\,[x+y_m(q_1)],
\label{eq:S1.4}
\end{equation}
where $x\equiv n_{2}/n_{1}$ is the local density ratio. The first term is just $\epsilon_{F2}$, the local Fermi energy for a noninteracting gas of atoms in state $2$. As the minority concentration vanishes, $x\rightarrow0$, we see that $\mu_{2}\rightarrow E_p(2)$, as expected. The interaction between the spin-components also modifies the chemical potential of the majority atoms in state $1$,
\begin{equation}
\mu_{1}=\epsilon_{F1}\,\left\{1+x\,[\,y_m(q_1)+y_m'(q_1)]\right\}.
\label{eq:S1.3}
\end{equation}
As $x\rightarrow0$, $\mu_{1}\rightarrow\epsilon_{F1}$ as it should. In the second term, we have defined $y_m'(q_1)\equiv dy_m(q_1)/d\log q_1=q_1 dy_m(q_1)/dq_1$,
\begin{equation}
y_m'(q_1)\equiv\frac{q_1[y_m(q_1)]^2}{1+2q_1}.
\end{equation}
The local 2D pressure is then $p=n_{1}\,\mu_{1}+n_{2}\,\mu_{2}-f$,
\begin{equation}
p=\frac{1}{2}\,n_{1}\,\epsilon_{F1}\left\{1+x^2+2x[\,y_m(q_1)+y_m'(q_1)]\right\}.
\label{eq:S2.4}
\end{equation}

To compare the predictions to the experimental measurements, we parameterize the interaction strength by
\begin{equation}
q_0\equiv \frac{E_{F}}{E_b},
\label{eq:S1.9}
\end{equation}
where $E_{F}\equiv\hbar\omega_\perp\sqrt{2N_1}$ is the ideal gas Fermi energy  for the majority component  at the center of a harmonic trap, with $\omega_\perp\equiv\sqrt{\omega_x\omega_y}$ the transverse oscillation frequency.
We define the ideal 2D gas unit of density
\begin{equation}
n_0\equiv\frac{E_{F}}{\alpha}=\frac{\hbar\omega_\perp}{\alpha}\sqrt{2N_1}=\frac{2}{\pi}\frac{N_1}{R_{TF1}^2},
\label{eq:S2.1}
\end{equation}
where $R_{TF1}\equiv\sqrt{2E_{F}/(m\omega_\perp^2)}$ is the Thomas-Fermi radius for majority component atoms of mass $m$.

The spatial profiles in the trapping potential are determined by the local chemical potentials, Eq.~\ref{eq:S1.4} and Eq.~\ref{eq:S1.3}. For simplicity, we assume a harmonic confining potential, where  $\mu_{i}=\mu_{i0}-m\omega_\perp^2\rho^2/2$, with $\rho$ the transverse radius. It is convenient to write the chemical potentials in units of $E_{F}$, $\tilde{\mu}_{i}=\mu_{i}/E_{F}$, the densities in units of $n_0$, $\tilde{n}_{i}\equiv n_{i}/n_0$, and the transverse radius in units of $R_{TF1}$, $\tilde{\rho}\equiv \rho/R_{TF1}$. Then, using $q_1=q_0\,\tilde{n}_{1}$,
\begin{equation}
\tilde{\mu}_{1}=\tilde{\mu}_{10}-\tilde{\rho}^2=\tilde{n}_{1}\,
\left\{1+x[\,y_m(q_0\,\tilde{n}_{1})+y_m'(q_0\,\tilde{n}_{1})]\right\}.
\label{eq:S4.1}
\end{equation}
\begin{equation}
\tilde{\mu}_{2}=\tilde{\mu}_{20}-\tilde{\rho}^2=\tilde{n}_{1}\,[x+y_m(q_0\,\tilde{n}_{1})].
\label{eq:S4.2}
\end{equation}
We solve these equations for the density profiles in two regions, $0\leq\tilde{\rho}<\tilde{R}_2$, where $\tilde{n}_{2}\neq 0$ and for $\tilde{R}_2\leq\tilde{\rho}\leq\tilde{R}_1$, where $\tilde{n}_{2}=0$. We begin with the latter.

For $\tilde{R}_2\leq\tilde{\rho}\leq\tilde{R}_1$, Eq.~\ref{eq:S4.1} with $x=0$ yields a Thomas-Fermi profile,
\begin{equation}
\tilde{n}_{1}(\tilde{\rho})=\tilde{\mu}_{10}-\tilde{\rho}^2.
\label{eq:S5.3}
\end{equation}
Since $\tilde{n}_{1}(\tilde{\rho})\geq 0$ for all $\tilde{\rho}$, we must have $\tilde{\mu}_{10}>0$ and $\tilde{\rho}^2\leq\tilde{\mu}_{10}$, which determines the radius at which the majority density vanishes,
\begin{equation}
\tilde{R}_1=\sqrt{\tilde{\mu}_{10}}.
\label{eq:S5.4}
 \end{equation}
 Further, for a gas with attractive interactions, $E_p(2)<0$, we expect $R_1$ to be smaller than the Thomas-Fermi radius, i.e., $\tilde{R}_1\leq 1$, which requires $0\leq\tilde{\mu}_{10}\leq 1$.

To find the radius at which the minority component vanishes, we set $x=0$ and $\tilde{\rho}=\tilde{R}_2$ in Eq.~\ref{eq:S4.2}. Then,
\begin{equation}
\tilde{R}_2^2=\tilde{\mu}_{20}-\tilde{n}_{1}(\tilde{R}_2)\,y_m[q_0\,\tilde{n}_{1}(\tilde{R}_2)],
\label{eq:S5.2}
\end{equation}
where $\tilde{n}_{1}(\tilde{R}_2)$ is known from Eq.~\ref{eq:S5.3} so that Eq.~\ref{eq:S5.2} can be solved numerically for $\tilde{R}_2$.

Given the chemical potentials at the trap center $\tilde{\mu}_{10},\tilde{\mu}_{20}$, the cutoff radii $\tilde{R}_1$
and $\tilde{R}_2$ are known. Then,  Eq.~\ref{eq:S4.1} and Eq.~\ref{eq:S4.2} can be solved for $\tilde{n}_{1}$ and $\tilde{n}_{2}$ in the region $0\leq\tilde{\rho}<\tilde{R}_2$. With $x\tilde{n}_{1}(\tilde{\rho})=\tilde{n}_{2}(\tilde{\rho})$, Eq.~\ref{eq:S4.2} immediately gives
\begin{equation}
\tilde{n}_{2}(\tilde{\rho})=\tilde{\mu}_{20}-\tilde{\rho}^2-\tilde{n}_{1}(\tilde{\rho})\,y_m[q_0\,\tilde{n}_{1}(\tilde{\rho})],
\label{eq:S6.1}
\end{equation}
where $\tilde{n}_{1}(\tilde{\rho})\equiv\tilde{n}_{1}$ is consistently determined for $0\leq\tilde{\rho}<\tilde{R}_2$ by eliminating $\tilde{n}_{2}(\tilde{\rho})$ from Eq.~\ref{eq:S4.1},
\begin{equation}
(\tilde{\mu}_{10}-\tilde{\rho}^2)-(\tilde{\mu}_{20}-\tilde{\rho}^2)[\,y_m(q_0\,\tilde{n}_{1})+y_m'(q_0\,\tilde{n}_{1})]=
\tilde{n}_{1}\,
\left\{1-y_m(q_0\,\tilde{n}_{1})[\,y_m(q_0\,\tilde{n}_{1})+y_m'(q_0\,\tilde{n}_{1})]\right\}.
\label{eq:S4.5}
\end{equation}

The density profiles are normalized according to $N_i=\int_0^\infty2\pi\rho\, d\rho\,n_{i}(\rho)$. For the majority, this requires
\begin{equation}
I_1=4\int_0^{\tilde{R}_1} \hspace{-0.1in}d\tilde{\rho}\,\tilde{\rho}\,\tilde{n}_{1}(\tilde{\rho})=1,
\label{eq:S6.3a}
\end{equation}
where $\rho=R_{TF1}\,\tilde{\rho}$ and we have used $n_{1}=n_0\,\tilde{n}_{1}$ with Eq.~\ref{eq:S2.1} for $n_0$.
For the minority, we require
\begin{equation}
I_2=4\int_0^{\tilde{R}_1} \hspace{-0.1in}d\tilde{\rho}\,\tilde{\rho}\,\tilde{n}_{2}(\tilde{\rho})=\frac{N_2}{N_1}.
\label{eq:S6.3b}
\end{equation}
In principle, the chemical potentials at the trap center $\tilde{\rho}=0$ can be determined from these normalization integrals. However, it is convenient to use the pressure at the trap center to constrain the chemical potentials for a given choice of the density ratio at the trap center, $0\leq x(0)\leq 1$.

The Gibbs-Duhem relation determines the pressure at the trap center. For fixed temperature,  $dp=n_{1}\,d\mu_{1}+n_{2}\,d\mu_{2}$. Since $d\mu_{1}=-dU_{trap}$, and the pressure vanishes for $U_{trap}\rightarrow\infty$, we have $$p(0)=-\int_\infty^0 [n_{1}(\rho)+n_{2}(\rho)]\,dU_{trap}.$$
For a harmonic trap,  $dU_{trap}=m\omega_\perp^2\rho\, d\rho$ and $U_{trap}\rightarrow\infty$ as $\rho\rightarrow\infty$, we immediately obtain
\begin{equation}
p(0)=\frac{m\omega_\perp^2}{2\pi}\int_0^\infty 2\pi\rho\,d\rho [n_{1}(\rho)+n_{2}(\rho)]=\frac{m\omega_\perp^2}{2\pi}(N_1+N_2).
\label{eq:S2.6b}
\end{equation}
This result is readily generalized for an anharmonic (gaussian) transverse trapping potential, as done in Ref.~\cite{PhysRevLett.112.045301}, leading to an additional negative term $\propto\langle\rho^2\rangle_i$ for each state $i=1,2$. Using Eq.~\ref{eq:S2.4}, we then have the desired constraint
\begin{equation}
\frac{1}{2}\left(1+\frac{N_2}{N_1}\right)=\frac{1}{2}\,\tilde{n}^2_{1}(0)\left\{1+x^2(0)+
2x(0)[\,y_m(q_0\,\tilde{n}_{1}(0))+y_m'(q_0\,\tilde{n}_{1}(0))]\right\}.
\label{eq:S3.2}
\end{equation}

Now we can find the density profiles. First, we select $q_0=E_{F}/E_b$ and $N_2/N_1$, i.e., the desired interaction strength and polarization, $P\equiv (N_1-N_2)/(N_1+N_2)$. Then, we choose a density ratio at the trap center $x(0)$ in the range $0\leq x(0)\leq 1$, which determines $\tilde{n}_{1}(0)$ from Eq.~\ref{eq:S3.2}. Together, $x(0)$ and $\tilde{n}_{1}(0)$ determine the chemical potentials at the trap center from Eqs.~\ref{eq:S4.1}~and~\ref{eq:S4.2},
\begin{eqnarray}
\tilde{\mu}_{10}&=&\tilde{n}_{1}(0)\,
\left\{1+x(0)[\,y_m(q_0\,\tilde{n}_{1}(0))+y_m'(q_0\,\tilde{n}_{1}(0))]\right\}\nonumber\\
\tilde{\mu}_{20}&=&\tilde{n}_{1}(0)\,[x(0)+y_m(q_0\,\tilde{n}_{1}(0))].
\end{eqnarray}
For each $x(0)$, we then find $\tilde{\mu}_{10}$ and $\tilde{\mu}_{20}$, which in turn determine possible density profiles. These are found as numerical interpolation functions for the regions  $0\leq\tilde{\rho}<\tilde{R}_2$ and $\tilde{R}_2\leq\tilde{\rho}<\tilde{R}_1$. Integrating the majority density, we will generally find that Eq.~\ref{eq:S6.3a} and Eq.~\ref{eq:S6.3b} yield $I_1\neq 1$, but $I_1+I_2=1+N_2/N_1$. By determining $I_1$ for  $x(0)$ in steps of $0.1$ and interpolating numerically, we find the value of $x(0))$ for which $I_1=1$, i.e., the $N_1$ integral is properly normalized. Then the $I_2$ integral yields $N_2/N_1$ as it should. With this value of $x(0)$, the 2D density profiles are determined for the given interaction strength and polarization.

Fig.~\ref{fig:radii} of the main text shows the measured radii for the minority $R_2$ and majority $R_1$ as a function of $N_2/N_1$, for $E_F/E_b=6.6$, $2.1$, and $0.75$. The polaron model, shown as the solid curves,  is in good agreement with the measurements. Here, we have extended the predictions for the imbalanced gas up to $N_2/N_1=0.9$. For $N_2/N_1=1$, we show the predictions for the balanced mixture, which are discussed below.

\begin{figure}[htb]
\includegraphics[width=5in]{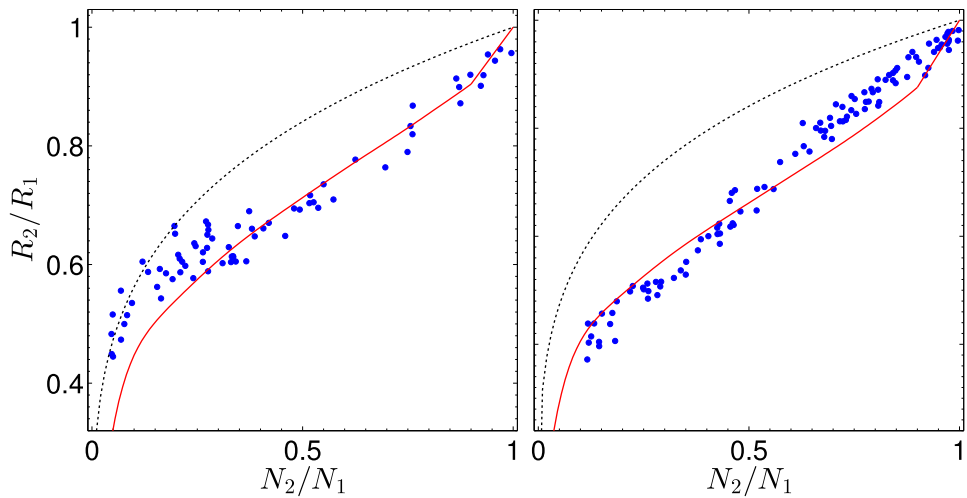}
\caption{Ratio of the minority to majority radius for versus $N_2/N_1$. Left: $E_{F}/E_b=2.1$;  Right: $E_{F}/E_b=0.75$.  Dots: Data; Solid line: 2D polaron model; Dashed line: Ideal Fermi gas prediction $(N_2/N_1)^{1/4}$.\label{fig:ratio}}
\end{figure}

Figs.~\ref{fig:ratio} shows the measured ratios $R_1/R_2$, which are nearly independent of $\omega_\perp$. The predictions for an ideal 2D Fermi gas, $(N_2/N_1)^{1/4}$, shown as dotted curves are compared to the predictions of the polaron model, shown as solid curves, which are in reasonable agreement with the data.
This agreement suggests that 2D polarons play an important role in determining the cloud profiles for the quasi-two-dimensional strongly interacting Fermi gas.

From the 2D density profiles, we find the column densities $n_{ci}(\tilde{x})$ in units of $N_1/R_{TF1}$, which are denoted $\tilde{n}_{ci}(\tilde{x})$, where $\tilde{x}=x/R_{TF1}$.  For the majority, in the interval $-\tilde{R}_1\leq\tilde{x}\leq\tilde{R}_1$,
\begin{equation}
\tilde{n}_{c1}(\tilde{x})=\frac{4}{\pi}\int_0^{\sqrt{\tilde{R}_1^2-\tilde{x}^2}}
\hspace{-0.2in}d\tilde{y}\,\tilde{n}_{1}(\sqrt{\tilde{x}^2+\tilde{y}^2}\,\,),
\label{eq:S7.1a}
\end{equation}
where the majority column density is normalized to 1.
For the minority, in the interval $-\tilde{R}_2\leq\tilde{x}\leq\tilde{R}_2$,
\begin{equation}
\tilde{n}_{c2}(\tilde{x})=\frac{4}{\pi}\int_0^{\sqrt{\tilde{R}_2^2-\tilde{x}^2}}
\hspace{-0.2in}d\tilde{y}\,\tilde{n}_{2}(\sqrt{\tilde{x}^2+\tilde{y}^2}\,\,).
\label{eq:S7.1b}
\end{equation}
Here, the minority column density is normalized to $N_2/N_1$.

\begin{figure}[htb]
\includegraphics[width=4.0in]{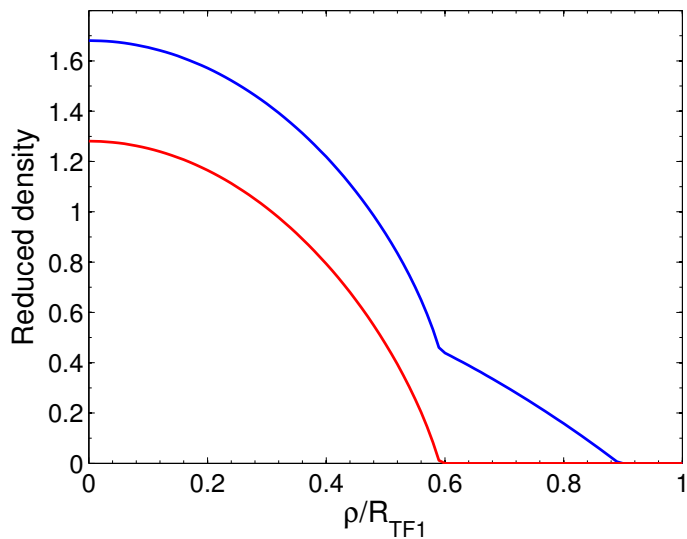}
\caption{Calculated 2D density profiles in units of $n_0$, Eq.~\ref{eq:S2.1}, for the minority (red lower) and majority (blue upper) for $E_{F}/E_b=0.75$ and $N_2/N_1=0.5$.  The interaction between the spin components significantly modifies the spatial profiles.\label{fig:imbaldensity}}
\end{figure}

Fig.~\ref{fig:imbaldensity} shows the calculated 2D density profiles for the majority and minority components for $E_{F}/E_b=0.75$ and $N_2/N_1=0.5$. For this case we find both majority and minority central densities greater than 1, and $x(0)=n_{2}(0)/n_{1}(0)=0.762$, $R_1=0.894\,R_{TF1}$ and $R_2=0.592\,R_{TF1}$.
In the region $R_2\leq\rho\leq R_1$, the majority density profile is of the Thomas-Fermi form. However, for $0\leq\rho <R_2$, the spatial profiles are strongly modified by attractive interactions between the two components.
For comparison, we note that for an ideal noninteracting Fermi gas, the corresponding 2D radius for the majority component is $R_{1ideal}=R_{TF1}$, while the 2D radius for the minority component is $R_{2ideal}=(N_2/N_1)^{1/4}R_{TF1}=0.841R_{TF1}$.

\begin{figure}[htb]
\includegraphics[width=5in]{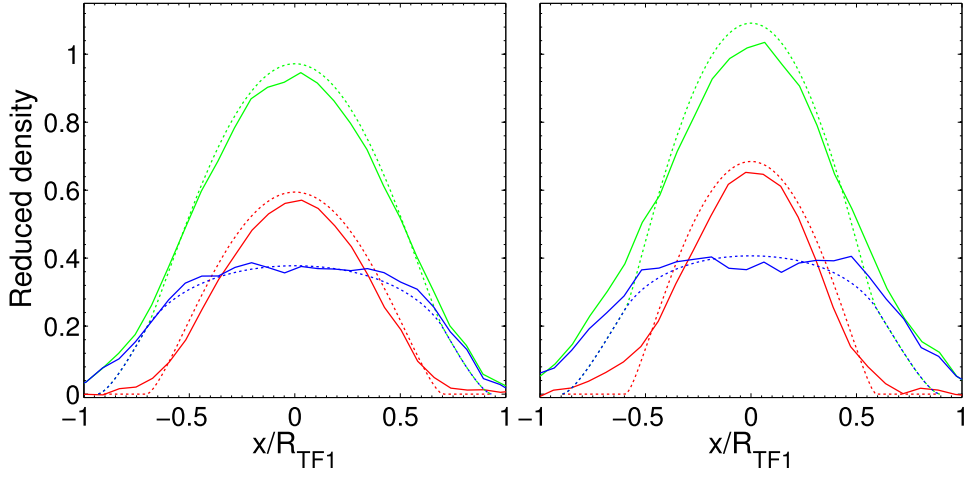}
\caption{Column density in units of $N_1/R_{TF1}$ for $N_2/N_1=0.5$. Left: $E_F/E_b=6.1$; Right: $E_F/E_b=0.75$. Green: 1-Majority; Red: 2-Minority. Blue: Density difference. Solid curves show the data; Dashed curves show the polaron model predictions for the same $\omega_\perp$ used throughout the paper, with no other adjustable parameters.\label{fig:1dprofiles}}
\end{figure}

The measured column densities for $N_2/N_1=0.5$ are shown in Fig.~\ref{fig:1dprofiles} as solid green and red curves. The difference in the measured column densities $n_{c1}-n_{c2}$ is shown as the solid blue curve. The dashed curves show the corresponding predictions based on Eq.~\ref{eq:S7.1a} and Eq.~\ref{eq:S7.1b}, with no adjustable parameters.
The predicted profiles for the column density difference, as well as for the individual profiles, are in reasonable agreement with the data, both in absolute peak density and width. However, for $E_F/E_b=0.75$, we see in Fig.~\ref{fig:1dprofiles} that the data for the difference in the column densities has a flatter profile and sharper edges than the polaron model prediction, consistent with a transition to a balanced  core.

\begin{figure}[!hb]
\includegraphics[width=4in]{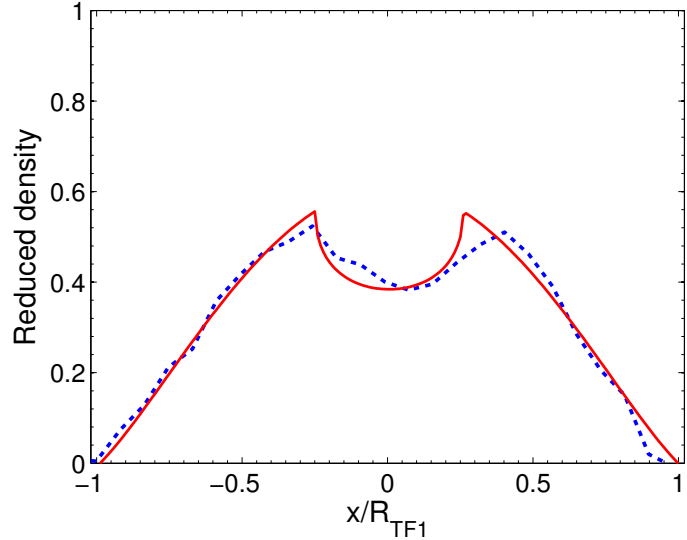}
\caption{Measured column density difference for $N_2/N_1=0.35$ at $E_F/E_b=0.75$ (dashed curve) compared to a balanced core model (red solid curve).\label{fig:balancedcore}}
\end{figure}

According to predictions of the polaron theory based on Eq.~\ref{eq:S1.6}, the 2D densities of two components never match if imbalanced. We now consider a model for the balanced core density profiles, which is consistent with our observations. For fully balanced 2D minority and majority distributions, we assume the minority density drops to zero beyond the balanced core of radius $R$ and the majority takes the form of a 2D Thomas-Fermi profile. The density difference is then $\Delta n_{2D}(\rho)=A\Theta[\rho-R]\Theta[R_1-\rho](1-\rho^2/R_1^2)$, where $\Theta$ is a Heaviside function. The corresponding column density difference, $\int dy\,\Delta n_{2D}(\sqrt{x^2+y^2})$ is fit to the data using $A$ and $R$ as free parameters. Fig.~\ref{fig:balancedcore} shows the result for $N_2/N_1 = 0.35$ and $E_F/E_b=0.75$,  demonstrating that a balanced core model is consistent with the measured column density profile.

\subsection{2D-Polaron Model: Spin-balanced mixtures.}
The density for the spin-balanced mixture is more easily obtained than for the imbalanced case.
For a spin-balanced mixture, $N_2=N_1$, we replace  Eq.~\ref{eq:Sfreeenergyimbal} with a free energy density that is symmetric in $n_{2}\leftrightarrow n_{1}$. Using the same notation as above, we take the polaron energies to be $E_p(2)=y_m(q_1)\,\epsilon_{F1}$ and $E_p(1)=y_m(q_2)\,\epsilon_{F2}$, so that
\begin{equation}
f=\frac{1}{2}\,n_{1}\,\epsilon_{F1}+\frac{1}{2}\,n_{2}\,\epsilon_{F2}
+\frac{1}{2}\,y_m(q_1)\,n_{2}\,\epsilon_{F1}+\frac{1}{2}\,y_m(q_2)\,n_{1}\,\epsilon_{F2}.
\label{eq:Sfreeenergybal}
\end{equation}
For the 3D problem at resonance, where $y_m=-0.6$, the same method with $1/2\rightarrow 3/5$, yields the correct pressure for the balanced gas. With the substitution, $\bar{\mu}\rightarrow (\mu_1+\mu_2)/2$, we recover the result of Ref.~\cite{PhysRevA.74.063628} for the balanced superfluid with a Bertsch parameter $1+y_m=0.4$.

From Eq.~\ref{eq:Sfreeenergybal}, we obtain the chemical potentials, $\mu_i=\partial f/\partial n_i$. Taking $n_2= n_1$
\begin{equation}
\mu_{1}=\mu_{2}=\epsilon_{F1}\,\left[1+\,y_m(q_1)+\frac{1}{2}y_m'(q_1)\right].
\label{eq:S6.4}
\end{equation}
Taking the total density to be $n$ so that $n_{1}=n_{2}=n/2$, we can write the free energy as
\begin{equation}
f=\frac{n}{2}\,\epsilon_{F1}[1+y_m(q_1)],
\label{eq:S7.1B}
\end{equation}
Here, we have used the same notation as above, with $\epsilon_{F1}=\alpha\,n_{1}$, $q_1\equiv q_0\,\tilde{n}_{1}$, and $\tilde{n}_{1}\equiv n_{1}/n_0$, where $n_0$ is given by Eq.~\ref{eq:S2.1}.
The corresponding local pressure, $p=n_{1}\mu_{1}+n_{2}\mu_{1}-f$, is then
\begin{equation}
p=\frac{n}{2}\,\epsilon_{F1}\,\left[1+y_m(q_1)+y_m'(q_1)\right].
\label{eq:S7.1C}
\end{equation}
For $E_b>>\epsilon_{F1}$, i.e., for $q\rightarrow 0$, Taylor expansion of $y_m$ and $y_m'$ shows that $1+y_m(q)+y_m'(q)\rightarrow 0$.

\begin{figure}[htb]
\includegraphics[width=4.0in]{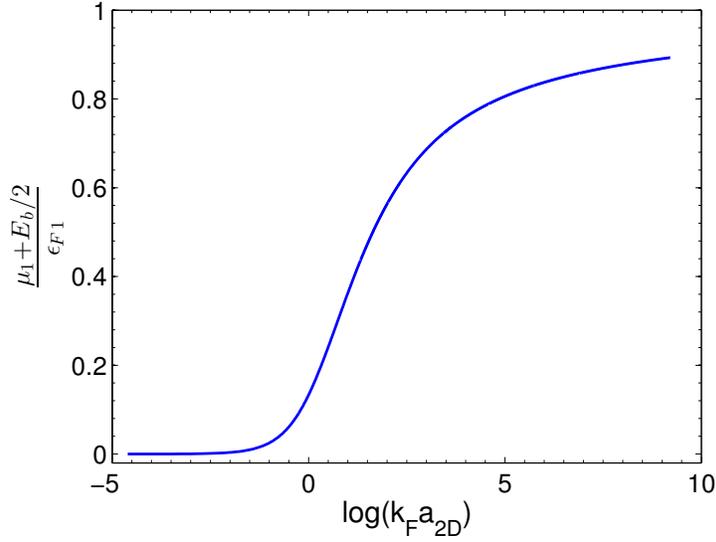}
\caption{2D Chemical potential from the polaron model for the balanced mixture versus the interaction parameter $log[k_F\,a_{2D}]$. Note that $\mu_1\rightarrow -E_b/2$ for $\epsilon_{F1}<<E_b$ as it should.\label{fig:balchempot}}
\end{figure}

Fig.~\ref{fig:balchempot} shows the chemical potential obtained from Eq.~\ref{eq:S6.4} in units of the local Fermi energy $\epsilon_{F1}$ as a function of $\log[k_F\,a_{2D}]$, where $k_F$ is the local Fermi wavevector, i.e., $k_F^2=4\pi\,n_1$ and $a_{2D}\equiv 2\,e^{-\gamma_E}\hbar/\sqrt{mE_b}$ is the 2D scattering length as defined in Ref.~\cite{1367-2630-13-11-113032}, where $\gamma_E=0.577$ is Euler's constant.
The chemical potential obtained in the polaron model using Eq.~\ref{eq:S1.6} for the balanced gas is in reasonable agreement with that of Fig.~1 of Ref.~\cite{1367-2630-13-11-113032}, which utilizes an interpolator adjusted to agree with the quantum Monte-Carlo predictions of Ref.~\cite{PhysRevLett.106.110403}.

Eq.~\ref{eq:S7.1C} can be used to determine the density at the trap center $\tilde{n}_{1}(0)$ and the corresponding chemical potential, by using the Gibbs-Duhem result, Eq.~\ref{eq:S2.6b} with $N_2=N_1$, $p(0)=m\omega_\perp^2\,N_1/\pi$. With Eq.~\ref{eq:S2.1}, the central density is then immediately determined by numerically solving
\begin{equation}
\tilde{n}_{1}(0)=\frac{1}{\sqrt{1+y_m[q_0\tilde{n}_{1}(0)]+y_m'[q_0\tilde{n}_{1}(0)]}}.
\label{eq:S10.1}
\end{equation}
We can write the chemical potential in units of $E_{F}$, using the same notation as in the previous section,
\begin{equation}
\tilde{\mu}_{1}=\tilde{n}_{1}\,\left[1+y_m(q_0\tilde{n}_{1})+\frac{1}{2}y_m'(q_0\tilde{n}_{1})\right].
\end{equation}
Using Eq.~\ref{eq:S10.1} for $\tilde{n}_{1}(0)$ then determines $\tilde{\mu}_{10}\equiv\tilde{\mu}_{1}(0)$. With $\tilde{\mu}_{1}=\tilde{\mu}_{10}-\tilde{\rho}^2$ as before, the density profile is then determined for the given $q_0=E_{F}/{E_b}$ using
\begin{equation}
\tilde{\mu}_{10}-\tilde{\rho}^2=
\tilde{n}_{1}(\tilde{\rho})\,
\left\{1+y_m[q_0\tilde{n}_{1}(\tilde{\rho})]+\frac{1}{2}y_m'[q_0\tilde{n}_{1}(\tilde{\rho})]\right\}.
\label{eq:S11.5}
\end{equation}

The density vanishes for $\rho >R_1$. To determine $R_1$, we consider Eq.~\ref{eq:S11.5} in the limit $\tilde{n}_{1}\rightarrow 0$, where Taylor expansion of $y_m$ and $y_m'$ shows that the  right-hand side approaches $-\tilde{n}_{1}/(2q_0)=-E_b/2$, which is half the dimer binding energy, as it should be for the chemical potential of an atom. With $\tilde{\rho}_{max}\equiv\tilde{R}_1$, the cloud radius in units of $R_{TF1}$ is then given by,
\begin{equation}
\tilde{R}_1=\sqrt{\tilde{\mu}_{1}(0)+\frac{1}{2q_0}}.
\label{eq:S12.3}
\end{equation}
We note that the density is then self-consistently normalized, i.e., it obeys Eq.~\ref{eq:S6.3a} as it should.
\begin{figure}[htb]
\includegraphics[width=4.0in]{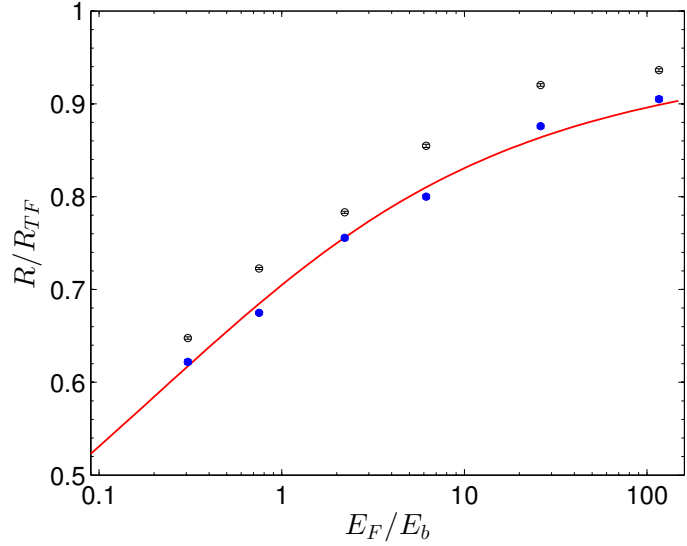}
\caption{Radius of the balanced mixture. Blue solid dots: Radius determined from Eq.~\ref{eq:R} using the measured central column density and total number $N_1$; Open black circles: Cut-off radius from ideal gas Thomas-Fermi fit; Red solid curve: Polaron model for the balanced mixture versus the interaction parameter $E_F/E_b$, where $E_F=\hbar\omega_\perp\sqrt{2N_1}$. \label{fig:balradius}}
\end{figure}

\subsection{Spin-balanced column density.}

The column density, Eq.~\ref{eq:S7.1a} for the balanced mixture is calculated by integrating the 2D spatial profiles obtained from  Eq.~\ref{eq:S11.5}. For a harmonic trap, we find that the predicted column density $n_{1D}(x)$ is very well fit by
\begin{equation}
n_{1D}(x)=n_{1D0}\left(1-\frac{x^2}{R^2}\right)^n\,\Theta[R-|x|],
\label{eq:1D}
\end{equation}
where $\Theta$ is a Heaviside function, $n_{1D0}$ is the peak column density, and $n$ is an exponent, determined from the fit. Normalizing the x-integral of Eq.~\ref{eq:1D} to the  number of atoms $N_1$ in one spin state yields,
\begin{equation}
R=\frac{N_1}{n_{1D0}\sqrt{\pi}}\frac{\Gamma(n+3/2)}{\Gamma(n+1)},
\label{eq:R}
\end{equation}
which determines $R$ from the measured atom number and peak column density, as shown in Fig.~\ref{fig:balradius} (blue solid dots). To find the peak column density we fit the
data by a parabola within 70\% of the apparent Thomas-Fermi radius. Thus, we avoid fitting the wings of the column density, which suffer from relatively high noise. For comparison, we also fit the measured profiles with the ideal gas Thomas-Fermi distribution to directly obtain the cut-off radii shown as open black circles in Fig.~\ref{fig:balradius}. Note, that the results of Thomas-Fermi fits are systematically higher, but can be fit very well by the polaron model over the whole range of the interaction strength $E_F/E_b$ by decreasing the transverse trap frequency by several per cent.

The corresponding 2D profile takes the form,
\begin{equation}
n_{2D}(\rho)=n_{2D0}\left(1-\frac{\rho^2}{R^2}\right)^{n-1/2}\Theta[R-\rho],
\label{eq:2D}
\end{equation}
where $\rho\equiv\sqrt{x^2+y^2}$. Normalization of Eq.~\ref{eq:2D} to the measured  number $N_1$ and elimination of $R$ using Eq.~\ref{eq:R}  relates  $n_{2D0}$ to $n_{1D0}$,
\begin{equation}
\tilde{n}_{2D0}=\frac{\pi}{2}\frac{[\Gamma(n+1)]^2}{\Gamma(n+1/2)\Gamma(n+3/2)}\,\tilde{n}_{1D0}^2.
\label{eq:2D1D}
\end{equation}
Here,  $\tilde{n}_{1D0}$ is the measured central column density in units of $N_1/R_{TF1}$ and $\tilde{n}_{2D0}$ is given in units of $n_0$, Eq.~\ref{eq:S2.1}. Thus, measuring the atom number and the peak column density is sufficient to extract the central 2D density without the need of fitting the wings of the column density profiles.

 The density profiles of Eqs.~\ref{eq:1D} and \ref{eq:2D} fit the spatial profiles predicted by the polaron model very well. For an ideal Fermi gas, we would have $n=3/2$ for the 1D fit. For the polaron model, we   find $n$ decreases as $q_0=E_F/E_{b}$ decreases, from $n=1.5$ at $q_0=100$, where the gas is nearly ideal, down to $n=1.03$ at $q_0=0.05$ where $E_b/E_F$ is large. Over this range, the first factor in Eq.~\ref{eq:2D1D} only varies from $1.39$ to $1.34$ and therefore is insensitive to $n$. Using the predicted power law exponents, Eq.~\ref{eq:2D1D} then relates $\tilde{n}_{2D0}$ to the measured $\tilde{n}_{1D0}$.

In the experiments, we determine the pressure at the trap center, in units of the total pressure $p_{ideal}$ of a spin-balanced ideal Fermi gas at the same density, as done in Ref.~\cite{PhysRevLett.112.045301}. As noted above, the Gibbs-Duhem relation, Eq.~\ref{eq:S2.6b}, yields the constant value $p(0)=m\omega_\perp^2\,N_1/\pi$ for a harmonic trap. Then, the reduced pressure for the balanced gas is determined simply from the 2D central density,
\begin{equation}
\tilde{p}(0)\equiv \frac{p(0)}{p_{ideal}}=\frac{1}{\tilde{n}^2_{1}(0)},
\label{eq:redpressexp}
\end{equation}
where $p_{ideal}=\epsilon_{F1}(n_1+n_2)/2=n_1\epsilon_{F1}$, and $\tilde{n}_{1}(0)$ is the 2D central density in ideal gas units $n_0$, Eq.~\ref{eq:S2.1}. We determine $\tilde{n}_{1}(0)\equiv\tilde{n}_{2D0}$ from the measured column density at the trap center, as discussed above. The measurements are compared with the polaron model predictions,  based on Eq.~\ref{eq:S7.1C},
\begin{equation}
\tilde{p}(0)=1+y_m[q_0\tilde{n}_{1}(0)]+y_m'[q_0\tilde{n}_{1}(0)],
\label{eq:redpressure}
\end{equation}
 where $\tilde{n}_{1}(0)$ is self-consistently determined for each $q_0$ using Eq.~\ref{eq:redpressexp}.

\begin{figure}[htb]
\includegraphics[width=4.0in]{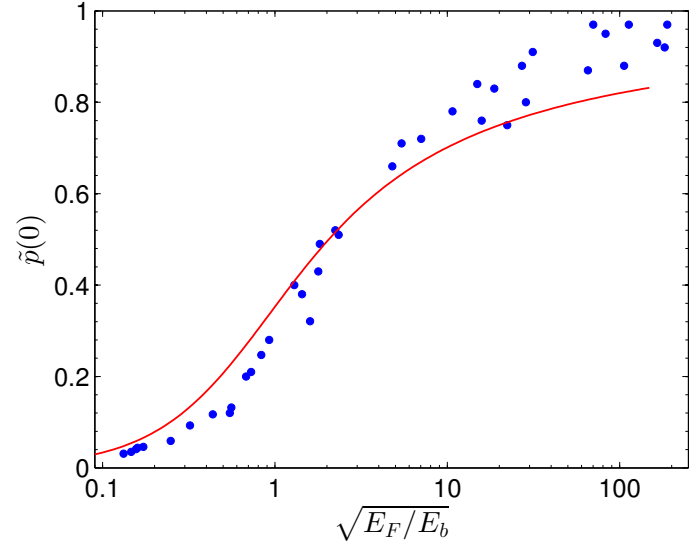}
\caption{Reduced 2D pressure $\tilde{p}(0)$ versus $E_F/E_b$, where $E_F=\hbar\omega_\perp\sqrt{2N_1}$, with $2N_1$ the total number of atoms. Dots: Data from Ref.~\cite{PhysRevLett.112.045301}. Solid curve: Prediction of Eq.~\ref{eq:redpressure}.  \label{fig:Andreypressure}}
\end{figure}

 Fig.~\ref{fig:pressure} of the main text shows the polaron prediction (solid curve) for the pressure of the balanced gas as a function of $E_F/E_{b}$, which agrees very well with the measurements.  Fig.~\ref{fig:Andreypressure} compares the predicted pressure with recent measurements over a larger range of $E_F/E_b$~\cite{PhysRevLett.112.045301}.
We obtain the curve shown and the best agreement in the small $E_b$ (BCS) limit by calculating $E_b$ for the unconfined case, $\omega_\perp/\omega_z=0$, using the parameters in the table of Ref~\cite{PhysRevLett.112.045301}. The polaron model is in reasonable agreement with the data and appears to reasonably approximate the quantum Monte-Carlo prediction of Ref.~\cite{PhysRevLett.106.110403}, which is shown in Ref~\cite{PhysRevLett.112.045301}.

We can compare the predicted pressure for the balanced gas with that of the imbalanced gas by replacing the central chemical potential for the balanced gas with the mean chemical potential obtained for the imbalanced gas. We find that the pressure of the balanced gas is always lower than that of the imbalanced gas, except when the densities of both components are equal, where the pressures match. Hence, for the simple polaron model based on Eq.~\ref{eq:S1.6}, we expect the gas to remain imbalanced for all $N_2/N_1<1$, in contrast to the observations.

\end{document}